

\message
{JNL.TEX version 0.92 as of 6/9/87.  Report bugs and problems to Doug Eardley.}

\catcode`@=11
\expandafter\ifx\csname inp@t\endcsname\relax\let\inp@t=\input
\def\input#1 {\expandafter\ifx\csname #1IsLoaded\endcsname\relax
\inp@t#1%
\expandafter\def\csname #1IsLoaded\endcsname{(#1 was previously loaded)}
\else\message{\csname #1IsLoaded\endcsname}\fi}\fi
\catcode`@=12



\font\twelverm=cmr10 scaled 1200    \font\twelvei=cmmi10 scaled 1200
\font\twelvesy=cmsy10 scaled 1200   \font\twelveex=cmex10 scaled 1200
\font\twelvebf=cmbx10 scaled 1200   \font\twelvesl=cmsl10 scaled 1200
\font\twelvett=cmtt10 scaled 1200   \font\twelveit=cmti10 scaled 1200
\font\twelvesc=cmcsc10 scaled 1200  \font\twelvesf=amssmc10 scaled 1200
\skewchar\twelvei='177   \skewchar\twelvesy='60


\def\twelvepoint{\normalbaselineskip=12.4pt plus 0.1pt minus 0.1pt
  \abovedisplayskip 12.4pt plus 3pt minus 9pt
  \belowdisplayskip 12.4pt plus 3pt minus 9pt
  \abovedisplayshortskip 0pt plus 3pt
  \belowdisplayshortskip 7.2pt plus 3pt minus 4pt
  \smallskipamount=3.6pt plus1.2pt minus1.2pt
  \medskipamount=7.2pt plus2.4pt minus2.4pt
  \bigskipamount=14.4pt plus4.8pt minus4.8pt
  \def\rm{\fam0\twelverm}          \def\it{\fam\itfam\twelveit}%
  \def\sl{\fam\slfam\twelvesl}     \def\bf{\fam\bffam\twelvebf}%
  \def\mit{\fam 1}                 \def\cal{\fam 2}%
  \def\sc{\twelvesc}		   \def\tt{\twelvett}
  \def\sf{\twelvesf}
  \textfont0=\twelverm   \scriptfont0=\tenrm   \scriptscriptfont0=\sevenrm
  \textfont1=\twelvei    \scriptfont1=\teni    \scriptscriptfont1=\seveni
  \textfont2=\twelvesy   \scriptfont2=\tensy   \scriptscriptfont2=\sevensy
  \textfont3=\twelveex   \scriptfont3=\twelveex  \scriptscriptfont3=\twelveex
  \textfont\itfam=\twelveit
  \textfont\slfam=\twelvesl
  \textfont\bffam=\twelvebf \scriptfont\bffam=\tenbf
  \scriptscriptfont\bffam=\sevenbf
  \normalbaselines\rm}



\def\beginlinemode{\endmode
  \begingroup\parskip=0pt \obeylines\def\\{\par}\def\endmode{\par\endgroup}}
\def\beginparmode{\endmode
  \begingroup \def\endmode{\par\endgroup}}
\let\endmode=\par
{\obeylines\gdef\
{}}
\def\singlespace{\baselineskip=\normalbaselineskip}

\def\oneandahalfspace{\baselineskip=\normalbaselineskip
  \multiply\baselineskip by 3 \divide\baselineskip by 2}
\def\doublespace{\baselineskip=\normalbaselineskip \multiply\baselineskip by 2}

\newcount\firstpageno
\firstpageno=2
\footline={\ifnum\pageno<\firstpageno{\hfil}\else{\hfil\twelverm\folio\hfil}\fi}
\def\toppageno{\global\footline={\hfil}\global\headline
  ={\ifnum\pageno<\firstpageno{\hfil}\else{\hfil\twelverm\folio\hfil}\fi}}
\let\rawfootnote=\footnote		
\def\footnote#1#2{{\rm\singlespace\parindent=0pt\parskip=0pt
  \rawfootnote{#1}{#2\hfill\vrule height 0pt depth 6pt width 0pt}}}
\def\raggedcenter{\leftskip=4em plus 12em \rightskip=\leftskip
  \parindent=0pt \parfillskip=0pt \spaceskip=.3333em \xspaceskip=.5em
  \pretolerance=9999 \tolerance=9999
  \hyphenpenalty=9999 \exhyphenpenalty=9999 }
\def\dateline{\rightline{\ifcase\month\or
  January\or February\or March\or April\or May\or June\or
  July\or August\or September\or October\or November\or December\fi
  \space\number\year}}
\def\received{\vskip 3pt plus 0.2fill
 \centerline{\sl (Received\space\ifcase\month\or
  January\or February\or March\or April\or May\or June\or
  July\or August\or September\or October\or November\or December\fi
  \qquad, \number\year)}}


\hsize=6.5truein
\hoffset=0truein
\vsize=8.9truein
\voffset=0truein
\parskip=\medskipamount
\def\\{\cr}
\twelvepoint		
\doublespace		
\overfullrule=0pt	




\def\title			
  {\null\vskip 3pt plus 0.2fill
   \beginlinemode \doublespace \raggedcenter \bf}

\def\author			
  {\vskip 3pt plus 0.2fill \beginlinemode
   \singlespace \raggedcenter\sc}

\def\affil			
  {\vskip 3pt plus 0.1fill \beginlinemode
   \oneandahalfspace \raggedcenter \sl}

\def\abstract			
  {\vskip 3pt plus 0.3fill \beginparmode
   \oneandahalfspace ABSTRACT: }

\def\endtitlepage		
  {\endpage			
   \body}
\let\endtopmatter=\endtitlepage

\def\body			
  {\beginparmode}		

\def\head#1{			
  \goodbreak\vskip 0.5truein	
  {\immediate\write16{#1}
   \raggedcenter \uppercase{#1}\par}
   \nobreak\vskip 0.25truein\nobreak}

\def\subhead#1{			
  \vskip 0.25truein		
  {\raggedcenter {#1} \par}
   \nobreak\vskip 0.25truein\nobreak}

\def\beginitems{
\par\medskip\bgroup\def\i##1 {\item{##1}}\def\ii##1 {\itemitem{##1}}
\leftskip=36pt\parskip=0pt}
\def\enditems{\par\egroup}

\def\beneathrel#1\under#2{\mathrel{\mathop{#2}\limits_{#1}}}

\def\refto#1{$^{#1}$}		

\def\references			
  {\head{References}		
   \beginparmode
   \frenchspacing \parindent=0pt \leftskip=1truecm
   \parskip=8pt plus 3pt \everypar{\hangindent=\parindent}}

\def\referencesnohead   	
  {                     	
   \beginparmode
   \frenchspacing \parindent=0pt \leftskip=1truecm
   \parskip=8pt plus 3pt \everypar{\hangindent=\parindent}}

\gdef\refis#1{\item{#1.\ }}			

\gdef\journal#1, #2, #3, 1#4#5#6{		
    {\sl #1~}{\bf #2}, #3 (1#4#5#6)}		

\def\pr{\journal Phys. Rev., }

\def\prb{\journal Phys. Rev. B, }

\def\prl{\journal Phys. Rev. Lett., }

\def\endreferences{\body}

\def\figurecaptions		
  {\endpage
   \beginparmode
   \head{Figure Captions}
}

\def\endpage			
  {\vfill\eject}

\def\endpaper			
  {\endmode\vfill\supereject}


\def\heading				
  {\vskip 0.5truein plus 0.1truein	
   \beginparmode \def\\{\par} \parskip=0pt \singlespace \raggedcenter}

\def\subheading				
  {\vskip 0.25truein plus 0.1truein	
   \beginlinemode \singlespace \parskip=0pt \def\\{\par}\raggedcenter}

\def\tag#1$${\eqno(#1)$$}

\def\align#1$${\eqalign{#1}$$}

\def\aligntag#1$${\gdef\tag##1\\{&(##1)\cr}\eqalignno{#1\\}$$
  \gdef\tag##1$${\eqno(##1)$$}}

\def\endaligntag{}

\def\overset #1\to#2{{\mathop{#2}\limits^{#1}}}
\def\underset#1\to#2{{\let\next=#1\mathpalette\undersetpalette#2}}
\def\undersetpalette#1#2{\vtop{\baselineskip0pt
\ialign{$\mathsurround=0pt #1\hfil##\hfil$\crcr#2\crcr\next\crcr}}}


\def\ref#1{Ref.~#1}			
\def\Ref#1{Ref.~#1}			
\def\[#1]{[\cite{#1}]}
\def\cite#1{{#1}}
\def\(#1){(\call{#1})}
\def\call#1{{#1}}
\def\taghead#1{}
\def\frac#1#2{{#1 \over #2}}

\def\12{{1\over2}}

\def\ie{{\it i.e.,\ }}

\def\etc{{\it etc.\ }}

\def\sla{\raise.15ex\hbox{$/$}\kern-.57em}
\def\leaderfill{\leaders\hbox to 1em{\hss.\hss}\hfill}
\def\twiddle{\lower.9ex\rlap{$\kern-.1em\scriptstyle\sim$}}
\def\bigtwiddle{\lower1.ex\rlap{$\sim$}}
\def\gtwid{\mathrel{\raise.3ex\hbox{$>$\kern-.75em\lower1ex\hbox{$\sim$}}}}
\def\ltwid{\mathrel{\raise.3ex\hbox{$<$\kern-.75em\lower1ex\hbox{$\sim$}}}}
\def\square{\kern1pt\vbox{\hrule height 1.2pt\hbox{\vrule width 1.2pt\hskip 3pt
   \vbox{\vskip 6pt}\hskip 3pt\vrule width 0.6pt}\hrule height 0.6pt}\kern1pt}
\def\tdot#1{\mathord{\mathop{#1}\limits^{\kern2pt\ldots}}}

\def\pmb#1{\setbox0=\hbox{#1}%
  \kern-.025em\copy0\kern-\wd0
  \kern  .05em\copy0\kern-\wd0
  \kern-.025em\raise.0433em\box0 }

\catcode`@=11
\newcount\r@fcount \r@fcount=0
\newcount\r@fcurr
\immediate\newwrite\reffile
\newif\ifr@ffile\r@ffilefalse
\def\w@rnwrite#1{\ifr@ffile\immediate\write\reffile{#1}\fi\message{#1}}

\def\writer@f#1>>{}
\def\referencefile{
  \r@ffiletrue\immediate\openout\reffile=\jobname.ref%
  \def\writer@f##1>>{\ifr@ffile\immediate\write\reffile%
    {\noexpand\refis{##1} = \csname r@fnum##1\endcsname = %
     \expandafter\expandafter\expandafter\strip@t\expandafter%
     \meaning\csname r@ftext\csname r@fnum##1\endcsname\endcsname}\fi}%
  \def\strip@t##1>>{}}

\def\citeall#1{\xdef#1##1{#1{\noexpand\cite{##1}}}}
\def\cite#1{\each@rg\citer@nge{#1}}	

\def\each@rg#1#2{{\let\thecsname=#1\expandafter\first@rg#2,\end,}}
\def\first@rg#1,{\thecsname{#1}\apply@rg}	
\def\apply@rg#1,{\ifx\end#1\let\next=\relax
\else,\thecsname{#1}\let\next=\apply@rg\fi\next}

\def\citer@nge#1{\citedor@nge#1-\end-}	
\def\citer@ngeat#1\end-{#1}
\def\citedor@nge#1-#2-{\ifx\end#2\r@featspace#1 
  \else\citel@@p{#1}{#2}\citer@ngeat\fi}	
\def\citel@@p#1#2{\ifnum#1>#2{\errmessage{Reference range #1-#2\space is bad.}%
    \errhelp{If you cite a series of references by the notation M-N, then M and
    N must be integers, and N must be greater than or equal to M.}}\else%
 {\count0=#1\count1=#2\advance\count1
by1\relax\expandafter\r@fcite\the\count0,%
  \loop\advance\count0 by1\relax
    \ifnum\count0<\count1,\expandafter\r@fcite\the\count0,%
  \repeat}\fi}

\def\r@featspace#1#2 {\r@fcite#1#2,}	
\def\r@fcite#1,{\ifuncit@d{#1}
    \newr@f{#1}%
    \expandafter\gdef\csname r@ftext\number\r@fcount\endcsname%
                     {\message{Reference #1 to be supplied.}%
                      \writer@f#1>>#1 to be supplied.\par}%
 \fi%
 \csname r@fnum#1\endcsname}
\def\ifuncit@d#1{\expandafter\ifx\csname r@fnum#1\endcsname\relax}%
\def\newr@f#1{\global\advance\r@fcount by1%
    \expandafter\xdef\csname r@fnum#1\endcsname{\number\r@fcount}}

\let\r@fis=\refis			
\def\refis#1#2#3\par{\ifuncit@d{#1}
   \newr@f{#1}%
   \w@rnwrite{Reference #1=\number\r@fcount\space is not cited up to now.}\fi%
  \expandafter\gdef\csname r@ftext\csname r@fnum#1\endcsname\endcsname%
  {\writer@f#1>>#2#3\par}}

\def\ignoreuncited{
   \def\refis##1##2##3\par{\ifuncit@d{##1}%
     \else\expandafter\gdef\csname r@ftext\csname
r@fnum##1\endcsname\endcsname%
     {\writer@f##1>>##2##3\par}\fi}}

\def\r@ferr{\endreferences\errmessage{I was expecting to see
\noexpand\endreferences before now;  I have inserted it here.}}
\let\r@ferences=\references
\def\references{\r@ferences\def\endmode{\r@ferr\par\endgroup}}

\let\endr@ferences=\endreferences
\def\endreferences{\r@fcurr=0
  {\loop\ifnum\r@fcurr<\r@fcount
    \advance\r@fcurr by 1\relax\expandafter\r@fis\expandafter{\number\r@fcurr}%
    \csname r@ftext\number\r@fcurr\endcsname%
  \repeat}\gdef\r@ferr{}\endr@ferences}


\let\r@fend=\endpaper\gdef\endpaper{\ifr@ffile
\immediate\write16{Cross References written on []\jobname.REF.}\fi\r@fend}

\catcode`@=12

\citeall\refto		
\citeall\ref		%
\citeall\Ref		%

\catcode`@=11
\newcount\tagnumber\tagnumber=0

\immediate\newwrite\eqnfile
\newif\if@qnfile\@qnfilefalse
\def\write@qn#1{}
\def\writenew@qn#1{}
\def\w@rnwrite#1{\write@qn{#1}\message{#1}}
\def\@rrwrite#1{\write@qn{#1}\errmessage{#1}}

\def\taghead#1{\gdef\t@ghead{#1}\global\tagnumber=0}
\def\t@ghead{}

\expandafter\def\csname @qnnum-3\endcsname
  {{\t@ghead\advance\tagnumber by -3\relax\number\tagnumber}}
\expandafter\def\csname @qnnum-2\endcsname
  {{\t@ghead\advance\tagnumber by -2\relax\number\tagnumber}}
\expandafter\def\csname @qnnum-1\endcsname
  {{\t@ghead\advance\tagnumber by -1\relax\number\tagnumber}}
\expandafter\def\csname @qnnum0\endcsname
  {\t@ghead\number\tagnumber}
\expandafter\def\csname @qnnum+1\endcsname
  {{\t@ghead\advance\tagnumber by 1\relax\number\tagnumber}}
\expandafter\def\csname @qnnum+2\endcsname
  {{\t@ghead\advance\tagnumber by 2\relax\number\tagnumber}}
\expandafter\def\csname @qnnum+3\endcsname
  {{\t@ghead\advance\tagnumber by 3\relax\number\tagnumber}}

\def\equationfile{%
  \@qnfiletrue\immediate\openout\eqnfile=\jobname.eqn%
  \def\write@qn##1{\if@qnfile\immediate\write\eqnfile{##1}\fi}
  \def\writenew@qn##1{\if@qnfile\immediate\write\eqnfile
    {\noexpand\tag{##1} = (\t@ghead\number\tagnumber)}\fi}
}

\def\callall#1{\xdef#1##1{#1{\noexpand\call{##1}}}}
\def\call#1{\each@rg\callr@nge{#1}}

\def\each@rg#1#2{{\let\thecsname=#1\expandafter\first@rg#2,\end,}}
\def\first@rg#1,{\thecsname{#1}\apply@rg}
\def\apply@rg#1,{\ifx\end#1\let\next=\relax%
\else,\thecsname{#1}\let\next=\apply@rg\fi\next}

\def\callr@nge#1{\calldor@nge#1-\end-}
\def\callr@ngeat#1\end-{#1}
\def\calldor@nge#1-#2-{\ifx\end#2\@qneatspace#1 %
  \else\calll@@p{#1}{#2}\callr@ngeat\fi}
\def\calll@@p#1#2{\ifnum#1>#2{\@rrwrite{Equation range #1-#2\space is bad.}
\errhelp{If you call a series of equations by the notation M-N, then M and
N must be integers, and N must be greater than or equal to M.}}\else%
 {\count0=#1\count1=#2\advance\count1
by1\relax\expandafter\@qncall\the\count0,%
  \loop\advance\count0 by1\relax%
    \ifnum\count0<\count1,\expandafter\@qncall\the\count0,%
  \repeat}\fi}

\def\@qneatspace#1#2 {\@qncall#1#2,}
\def\@qncall#1,{\ifunc@lled{#1}{\def\next{#1}\ifx\next\empty\else
  \w@rnwrite{Equation number \noexpand\(>>#1<<) has not been defined yet.}
  >>#1<<\fi}\else\csname @qnnum#1\endcsname\fi}

\let\eqnono=\eqno
\def\eqno(#1){\tag#1}
\def\tag#1$${\eqnono(\displayt@g#1 )$$}

\def\aligntag#1\endaligntag
  $${\gdef\tag##1\\{&(##1 )\cr}\eqalignno{#1\\}$$
  \gdef\tag##1$${\eqnono(\displayt@g##1 )$$}}

\def\eqalignno#1{\displ@y \tabskip\centering
  \halign to\displaywidth{\hfil$\displaystyle{##}$\tabskip\z@skip
    &$\displaystyle{{}##}$\hfil\tabskip\centering
    &\llap{$\displayt@gpar##$}\tabskip\z@skip\crcr
    #1\crcr}}

\def\displayt@gpar(#1){(\displayt@g#1 )}

\def\displayt@g#1 {\rm\ifunc@lled{#1}\global\advance\tagnumber by1
        {\def\next{#1}\ifx\next\empty\else\expandafter
        \xdef\csname @qnnum#1\endcsname{\t@ghead\number\tagnumber}\fi}%
  \writenew@qn{#1}\t@ghead\number\tagnumber\else
        {\edef\next{\t@ghead\number\tagnumber}%
        \expandafter\ifx\csname @qnnum#1\endcsname\next\else
        \w@rnwrite{Equation \noexpand\tag{#1} is a duplicate number.}\fi}%
  \csname @qnnum#1\endcsname\fi}

\def\ifunc@lled#1{\expandafter\ifx\csname @qnnum#1\endcsname\relax}

\let\@qnend=\end\gdef\end{\if@qnfile
\immediate\write16{Equation numbers written on []\jobname.EQN.}\fi\@qnend}

\catcode`@=12


\def\ie{{\it i.e.,\ }}

\def\etc{{\it etc.\ }}

\def\>{\rangle}
\def\<{\langle}
\def\o{\over}

\def\prop{\propto}

\def\slD{\raise.15ex\hbox{$/$}\kern-.57em\hbox{$D$}}
\def\dsl{\raise.15ex\hbox{$/$}\kern-.57em\hbox{$\Delta$}}
\def\slp{{\raise.15ex\hbox{$/$}\kern-.57em\hbox{$\partial$}}}
\def\nsl{\raise.15ex\hbox{$/$}\kern-.57em\hbox{$\nabla$}}
\def\sla{\raise.15ex\hbox{$/$}\kern-.57em\hbox{$\rightarrow$}}
\def\slla{\raise.15ex\hbox{$/$}\kern-.57em\hbox{$\lambda$}}
\def\slb{\raise.15ex\hbox{$/$}\kern-.57em\hbox{$b$}}
\def\lnp{\raise.15ex\hbox{$/$}\kern-.57em\hbox{$p$}}
\def\lnk{\raise.15ex\hbox{$/$}\kern-.57em\hbox{$k$}}
\def\lnK{\raise.15ex\hbox{$/$}\kern-.57em\hbox{$K$}}
\def\lnq{\raise.15ex\hbox{$/$}\kern-.57em\hbox{$q$}}

\def\a{\alpha}

\def\ga{{\gamma}}
\def\de{{\delta}}
\def\eps{{\epsilon}}

\def\th{{\theta}}

\def\la{\lambda}

\def\om{{\omega}}
\def\Ga{{\Gamma}}

\def\De{{\Delta}}

\def\munu{{\mu\nu}}

\def\cL{{\cal L}}

\def\part{\partial}

\def\dag{\dagger}

\def\abs{
         \vskip 3pt plus 0.3fill\beginparmode
         \doublespace ABSTRACT:\ }

\singlespace
\def\skiip#1{}
\def\veps{\varepsilon}

\title
Sideways Tunnelling and Fractional Josephson
Frequency in Double Layered Quantum Hall Systems

\author
X. G. Wen$^1$ \& A. Zee $^{2, 3}$

\affil
$^1$ Department of Physics
Massachusetts Institute of Technology
77 Massachusetts Avenue
Cambridge, MA 02139, USA

\affil
$^2$ Physique Theorique
Ecole Normale Superieure
75231 Paris,  France
\affil
$^3$ Institute for Theoretical Physics
University of California
Santa Barbara, CA 93106, USA

\abs{
The $(kkk)$ double layered quantum Hall state has been shown to exhibit many
superfluid properties including the appearance of an alternating tunnelling
current in response to an external voltage V. Here we propose the notion of
``sideways tunnelling", in which a current tunnels between two such
$(kkk)$ states through an incompressible fractional
quantum Hall state $(mm0)$. We predict the tunnelling current will alternate
with
a frequency $\om={eV\o \hbar m}$, as a consequence of the tunnelling of
$e/m$ fractional charges. We also discuss singularities in
the tunnelling current noise
spectrum and the non-linear I-V curve.
}
\endtopmatter

Recently we predicted,\refto{WZ}
within the semiclassical approximation and in
the weak tunnelling limit, that a DC voltage $V$ applied
across a double-layered quantum Hall system\refto{exp} in
certain special states (to be characterized below)
would produce a tunnelling current alternating with
frequency $\omega = eV/ \hbar$.
While our effect is reminiscent of the Josephson effect\refto{JS} there are
some
important differences. Here the
frequency is half of the Josephson frequency: the charge carrier is not
a pair of electrons as in a superconductor.\refto{wil}
We claimed that there is a superfluid lurking in certain
states of double-layered quantum Hall systems.
In this paper, we propose and discuss a novel notion of ``sideways tunnelling"
involving the juxtaposition of three double layered quantum Hall
systems.
The discussion here is orthogonal to that in \Ref{WZ}: in particular,
we will neglect interlayer tunnelling\refto{EW} in the main part of this paper.

In the effective Chern-Simons gauge field theory\refto{WZ2}
description of quantum Hall fluids, a double layered system is
described by
$$
\cL={1\over 4\pi}(\sum_{I,J}\veps_{\mu\nu\la}
a_{I\mu} K_{IJ}\part_\nu a_{J\la}
+2\sum_{I}\veps_{\mu\nu\la}A_\mu\part_\nu a_{I\la})
    + {\rm Maxwell\ terms}  \eqno(2.1)
$$
Here $I,J=1,2$ and $K$ is a $2\times 2$ matrix.
The electromagnetic current in layer $I$ is given by
$ J_\mu^{(I)}={1\over2\pi}\eps^{\munu\la}\part_\nu a_{I\la} $
with the total density and current
$J_\mu=J^{(1)}_\mu+J_\mu^{(2)}$ coupling to the
electromagnetic potential $A_\mu$. The difference
$J_\mu^{(-)}=J^{(1)}_\mu-J_\mu^{(2)}$ describes the relative density and
current fluctuations in the two layers.

When the matrix $K$ has a zero eigenvalue, the corresponding
eigenmode describes a superfluid.
For the sake of simplicity and
definiteness we specialize to the matrices $K =  \pmatrix{ k&k\cr k&k\cr}$
with $k$ an odd integer.
(This describes the class of states studied in \Ref{WZ}
and corresponds to the wavefunction $(k k k)$ in Halperin's
notation\refto{BH}.) The mode $a_- = a_1 -a_2$ is then governed by Maxwell
dynamics (which is gapless\refto{MAC}), rather than Chern-Simons dynamics, and
thus describes a superfluid.

In our formalism, interlayer tunnelling events correspond to the presence of
magnetic monopoles in space-time.
When an electron tunnels from one layer to the other
$$
\pm 2= \int d^2xdt\ \part^\mu J^{(-)}_{\mu}=\int d^2xdt\ {1\over 2\pi}
\part^\mu\eps_{\munu\la}\part_\nu\a_-^\la
\eqno(2.4)
$$
The discrete character of the electron implies Dirac quantization of the
magnetic monopole
associated with the gauge potential $a_-$ in Eculidean 3-space. Going
though the analysis of \ref{WZ}, we found that an applied voltage
produces an alternating current of frequency $\om=eV/\hbar$.
Note that the frequency does not depend on $k$.
This is because only electron can tunnel from
one layer to the other.  The vacuum separating the
layers does not support quasiparticle
excitations and hence forbids quasiparticle tunnelling.
Thus the frequency of the interlayer tunnelling
does not reflect the fractional charge of the quasiparticles.
Mathematically, the potential $a_-$ corresponds to the zero eigenvalue of $K$
and thus does not ``know" about $k$.

How does the long distance effective theory recognize the selection rule
that only electrons can tunnel? When
a charge $e/k$ quasiparticle tunnels, we have, by comparison with \(2.4), a
magnetic monopole with $2/k$ the Dirac magnetic charge, which we may
call a fractional monopole.
Dirac quantization implies that the fractional monopole has attached to
it a flux tube or string which carries  $ 2/k$ units of flux and
which ends on a fractional antimonopole. This flux string cannot be gauged
away, because the charges that couple to the $a_-$ are quantized as integers.
(Those charges represent vortex excitations in the Hall fluid.\refto{WZ2})
 The fractional monopole and the
fractional antimonopole are thus linearly confined by the string tension.
Physically, this means
that after a quasiparticle tunnels from one layer to another,
within a short time determined by the confinement length it
will tunnel back. Quasiparticle tunnelling cannot be observed at low
energies.

To have more precise description of tunnelling,
let us consider the ``noise" spectrum defined by
$$
F(\om)=\int^{\infty}_0 dt\ \<j_T(t) j_T(0)\> e^{-i\om t}
\eqno(2w)$$
with $j_T$ the interlayer tunnelling current
$$j_t=i\int d^2x\ [t(x)c^\dag_1(x) c_2(x) -h.c.].
\eqno(a7)$$
(where $c_i$ represents an electron annihilation operator in layer $i$.)
An ideal Josephson effect is defined by
a $\de$-function in $F(\om)$ at the Josephson frequency.
This occurs for interlayer tunnelling at zero temperature
(within perturbation theory in the tunnelling amplitude $t$).\refto{WZ}

So far we have reviewed the situation discussed in \ref{WZ}. Let us now
come to what we call ``sideways tunnelling".
{}From now on we will assume for simplicity
that there is no interlayer tunnelling. We imagine
an experimental situation in which three double-layered
systems characterized by matrices $K_L$, $K_M$,
and $K_R$ (with L, M, R for left, middle, and right respectively)
are connected together as shown in figure 1.
For definiteness
let us take $K_L$ and $K_R$ to be $\pmatrix{k&k\cr k&k\cr}$ and $K_M$ to be
$\pmatrix{m&0\cr 0&m\cr}$ with $m$ an odd integer.
We apply a voltage $V_R$ between the top and bottom layers
in the R region and $V_L$ in the L region.

Consider a quasiparticle of charge $e/m$
tunnelling sideways from the left edge to the right edge of the middle region
in
the top layer, and at the same time
a quasihole tunnelling in the same direction in
the bottom layer.
The coupling between the order parameter $\<c^\dag_1 c_2\>$ in the R region
and the excitations on the right edge of the M region is induced by
an electron tunnelling across the boundary in the top layer and a
hole tunnelling in the bottom layer. A similar coupling exists at the boundary
between the L and M region. This would be made more precise below (see
eqs. (9,10)).  Assuming
equilibrium between the bulk of the R, L regions and the  two edges of the
M region, we see that the voltage difference between the top and bottom
layers at the right edge of the M region is given by $V_R$ and at the left edge
by $V_L$. Thus the quasiparticle
tunnelling process described above costs an
energy $e(V_R-V_L)/m$. If  the tunnelling is coherent,
the frequency of the alternating current will be
$$
\omega = {\De E \over \hbar} = {1\o m} e|V_L - V_R|/\hbar\equiv \om_m
\eqno(ad1) $$
In general the tunnelling noise spectrum may also have peaks at multiples of
the above frequency.
The peak at $\om = m\om_m$ is due to electron
tunnelling. Since the long range orders exist in the R and L regions,
this peak
is a $\de$-function peak (at zero temperature).

If $k\neq m$, then by charge conservation,
quasiparticle tunnelling necessarily involves gapless edge excitations
between the L, M and and between the R, M regions.
We will see later that the peak at $\om= \om_m$
can be an algebraic singularity (due to interferences with the edge
excitations),
 or a $\de$-function singularity, depending on the strength of
the coupling to edge excitations.

The sideways tunnelling moves charge $e/m$ across the middle region in the
first
layer and charge $-e/m$ in the second layer. Thus the sideways tunnelling is
represented by a {\it pair} of fractional monopole and antimonopole
across the M region.
This dipole and the reversed dipole do not confine. Hence the tunnelling and
reversed tunnelling are not correlated in time and
do not cancel each other. This suggests
that, in contrast to the interlayer tunnelling,
sideways tunnelling of quasiparticles can be observed at low energies.

Edge excitations
on the two sides of M region carry different momenta.
To conserve momentum in the tunnelling process,
impurities must intervene: in other words,
an excitation going from the left edge to the right edge
must scatter off and exchange momentum with impurities.
For simplicity first ignore the R and L regions.
Quasiparticle tunnelling between the two edges assisted by a single impurity
already contains an algebraic singularity at a fractional Josephson
frequency in its noise spectrum.\refto{edge}
But in general there are many impurities, impurities a, b, c, \etc,
with effective voltages across each impurity
(\ie $V_R-V_L$ near each impurity) $V_a \neq V_b......$
and so the algebraic singularity in the $F(\om)$
would be completely smeared.
At this point, the hidden superfluid comes to the rescue.
For the $K_L $ and $ K_R $ we chose,
there is a superfluid in the left and right
regions. In the superfluid the voltage difference between the two layers are
constant in space (\ie $V_L$ and $V_R$ are constant in the region $R$ and $L$).
The superfluidity insures that all the effective
voltages $V_a, V_b,    ....$ are equal,
and hence guarantees the existence of,
at least,
an algebraic singularity at $\om=\om_m$.

To gain a better understanding of sideways tunnelling and the effects of
edge excitations, let us give a brief description
of the relevant edge states. For our choice of $K_{L,R,M}$,
there are three branches of edge excitations on the borders between
L, M regions and between the R, M regions. The $K_R$ state in the R
region has one branch of edge excitations. Let $\psi_{eR}$ denote the electron
operator in that edge branch. The electron creation operators in the $1^{st}$
and $2^{nd}$
layers are given by
$$c_{R1}^\dag=\psi_{eR}^\dag e^{-i\th_R/2},\ \ \ \
c_{R2}^\dag=\psi_{eR}^\dag e^{i\th_R/2}
\eqno(15)$$
at the edge of the R region, where $\th_R$ is the phase of the superfluid
order parameter in the R region. The $K_M$ state contains
two branches of edge excitations associated with the top and the
bottom layers. Let $\chi_{qRi}$ to be the quasiparticle operator in the
$i^{th}$ layer and at the right edge of the middle region. The operators
$\chi_{eRi}=(\chi_{qRi})^m$, carrying charge $e$, are the corresponding
electron  operators.
Similarly we can define $\psi_{eL}$, $c_{Lj}$, $\chi_{qLj}$, and $\chi_{eLj}$.

Now the coupling between the R, M regions
can be described by the following operator (and its hermitian conjugate)
$$\eqalign{
A_{RM}=&\int dx\ t_R( c_{R1}^\dag \chi_{eR1}+c_{R2}^\dag \chi_{eR2}
        )   \cr
}
\eqno(a8)$$
induced by electron tunnelling. Because of the superfluid
correlation $\<c_1^\dag c_2\>\neq 0$, $A_{RM}$ induce
an effective interlayer
coupling between $\chi_{eR1}$ and $\chi_{eR2}$:
$$\eqalign{
B_{RM}=& \int dx\ t'_R \chi_{eR2}^\dag \chi_{eR1} e^{-i\th_R}\cr
}
\eqno(a1)$$
Similarly, $A_{LM}$ and $B_{LM}$ are defined by substituting everywhere
L for R.
The quasiparticle tunnelling between the two edges of the
middle region described above is given by
$$A_M=\int dx\ t_M \chi_{qR1}\chi_{qR2}^\dag\chi_{qL1}^\dag\chi_{qL2}
$$
In general $t_{R,L}$, $t'_{R,L}$, and $t_M$ depend on $x$ due to impurities.
Note that $A_M$ refer only to the M region, while $A_{RM,LM}$ and $B_{RM,LM}$
link the M region with the R and L regions.
The tunnelling current across the middle region
is given by
$$I_{-}(t)=\int^t dt' (e^{i{e\o m}(V_L-V_R)t'} \<[A^\dag_M(t'), A_M(0)]\>+c.c.
)
\eqno(tt)$$
(Here we have assumed
$t_M \ll t_R, t_L$ and the equilibrium between the edge and the bulk.)
Note $I_{-}$ is the difference between the tunnelling currents in the top
and the bottom layer $I_-=I_1-I_2$. The total tunnelling current $I_1+I_2$
vanishes for the tunnelling described by $H_M$.
In the following we will assume that $t_{R,L,M}$ have random phases and short
range correlation so that $\<t_R(x)t_R(y)\> \sim \de(x-y)$ \etc.

We will now determine the long time behavior of the correlation between
$A_M(t)$
and $A_M(0)$, and hence by \(tt) the frequency characteristic of the current
$I_-$. We start with the long time behavior between various relevant operators
known from the theory of the edge excitations\refto{edge}
(a schematic notation suffices for our purpose):
$$\eqalign{
\< \chi_q^\dag(t) \chi_q(0)\> \sim & {1\o t^{1/m}} \cr
\< \chi_e^\dag(t) \chi_e(0)\> \sim & {1\o t^{m}} \cr
\< \psi_e^\dag(t) \psi_e(0)\> \sim & {1\o t^{k}} \cr}
\eqno(14)$$
Using these known facts, we now determine the scaling dimensions of
various operators.
After averaging \(a8) over the locations of impurities,
we induce in the action the term
$\de S\propto \int dt dt' dx \ c^\dag_{R1}(x,t)\chi_{eR1}(x,t)
c_{R1}(x,t')\chi^\dag_{eR1}(x,t')$.
 From \(14) and \(15),
$\<(c_{Rj}^\dag \chi_{eRj})_{x,t}(c_{Rj} \chi_{eRj}^{\dag })_0\>
\prop (x^2-v^2t^2)^{-(k+m)/2}$.
Thus $\de S$ and hence $A_{RM}$ (and $A_{LM}$) are irrelevant if $k+m-3>0 $.
Similarly,
we can show that $B_{RM}$ (and $B_{LM}$) are irrelevant if $2m-3>0$.

Two situations need to be discussed separately. When the two above
inequalities are satisfied, and if $t_{R,L}$ are small,
 the correlation function of $A_M$ can be
calculated neglecting $A_{RM,LM}$ and $B_{RM,LM}$.
We find, averaging over $t_M$, that $\<[A^\dag_M(t), A_M(0)]\>\prop
L_{edge}t^{-4/m}$, where $L_{edge}$ denotes the length of the edge.
Thus, in the weak coupling limit, the I-V curve has the form
$I_-\prop (V_L-V_R)^{{4\o m}-1}$
and the singularity  in the noise spectrum $F(\om)$ is given by
$$F(\om)\prop {1\o \Ga(4/m)} (\om-\om_m)^{{4\o m}-1}
\eqno(a3)$$
In this discussion we have ignored the potential interaction between
the edge branches. This potential interaction may change the above exponent,
 \ie one should replace $4/m$ by a new exponent $\ga$ whose value
depends on the interaction.

When $t_{R,L}$ are large and/or one of $A_{RM,LM}$ and $B_{RM,LM}$ is relevant,
 $B_{RM,LM}$ are important at
low energies.
This coupling not only makes $\<\chi_{eR2}^\dag \chi_{eR1}\>\neq 0$,
it also makes $\<\chi_{qR2}^\dag \chi_{qR1}\>\neq 0$.
The latter implies the ideal Josephson effect (at $T=0$), \ie
a finite $I_-$ even when $V_R-V_L=0$ and a $\de$-function singularity
in $F(\om)$ at frequency $\om=\om_m$ determined by the
fractional charge.
When $\<\chi_{qR2}^\dag \chi_{qR1}\>\neq 0$, its phase is not uniquely
determined by $\th_R$. The system chooses one of $m$ different ground states
characterized by $\<\chi_{qR2}^\dag \chi_{qR1}\>\propto
e^{i(\th_R+2n\pi)/ m}$, where $n=0,...,m-1$.

To show that the coupling \(a1) can indeed induce a non-zero
$\<\chi_{qR2}^\dag \chi_{qR1}\>$,
take $t'_R$ in \(a1) proportional to a $\de$-function. The tunnelling
events induced by \(a1) can be described by a 1D gas (in time direction) in
the path integral formalism. The long time
correlation
$\<(\chi_{eR2}^\dag \chi_{eR1})_t (\chi_{eR1}^\dag \chi_{eR2})_0 \>
\sim {1\over t^{2m}} \sim e^{-2m \log t}$
can be intepreted as
logarithmic interaction $-2m\ln|t_1-t_2|$ between the
particles in the Coulomb gas. The properties of such 1D Coulomb
gas has been studied in \Ref{cg}.

The Coulomb gas, in general, can have two phases if $m>1$:
the plasma
phase (if $t'_R$ is large) and the confining phase (if $t'_R$ is small).
If $m\leq 1$ the Coulomb gas is always in the plasma phase.
(In the present context, $m$ cannot be less than 1 and the $m = 1$
case does not involve tunnelling of fractional charges.)

In the plasma phase positive and
negative charges move freely and completely screen any added charges.
Insertion of the operator $\chi_{qR2}^\dag \chi_{qR1}$ is just like adding
charges to the Coulomb gas. Thus
the correlation  of $\chi_{qR2}^\dag \chi_{qR1}$ in the time direction
can be written as
$$ \< (\chi_{qR2}^\dag \chi_{qR1})_t (\chi_{qR1}^\dag \chi_{qR2})_0 \>
\prop  e^{-V(t)}
\eqno(a4)$$
where $V(t)$ is the potential between the added charges.
With $V$ short ranged, $\chi_{qR2}^\dag \chi_{qR1}$ has a long
range correlation.

For small $t'_R$ and $m>1$, the gas is in the confining phase
with the positive and negative charges forming a bound state, which
can only
modify the strength of the logarithmic interaction between the added charges.
With $V$ logarithmic, $\chi_{qR2}^\dag \chi_{qR1}$ decays algebraically,
thus implying an algebraic singularity in the noise spectrum $F(\om)$.

We have shown that for small $t'_R$ and at $T=0$,
the noise spectrum
$F(\om)$ contains an algebraic singularity at $\om=\om_m$. At finite
temperatures this algebraic singularity is expected to be smeared to have a
finite width of order $T$. For large $t_R$ and at $T=0$,
the noise spectrum $F(\om)$ contains a $\de$-function peak. At
finite temperatures the $\de$-function singularity is smeared to, at most,
an algebraic singularity.
This is
because the order parameters in the R,L regions have an algebraic  decay
at finite temperatures.

In the presence of interlayer tunnelling, all the algebraic and
$\de$-function singularities in the sideways tunnelling discussed here are
expected
to be modified below an energy scale $\De$ proportional to the square root of
the interlayer tunnelling amplitude,\refto{MAC,WZ} an energy scale describing
the gap for the relative density fluctuations between the two layers.
They may be smeared into peaks with finite width $\De$ since the $U(1)$
symmetry is explicitly broken by the interlayer tunnelling.
However, as long as $\De$ is small we would  still expect to see
narrow band noises at multiples of $\om_m$.

This work is supported by the National Science
Foundation under Grant No. DMR91-14553 (for XGW) and PHY89-04035 (for AZ).
One of us (AZ) also thanks the Ecole Normale Superieure for hospitality and
support.

\references

\refis{WZ}  X. G. Wen, and A. Zee,
\prl 69, 1811, 1992, \pr B47, 2265,  1993.

\refis{EW}    For an attempt to generalize \Ref{WZ} to the three-layer case,
see F. Ezawa , A. Iwazaki, and Y. S. Wu, Sendai preprint TU-421.

\refis{WZ2} X. G. Wen, and A. Zee, \pr  B46, 2290, 1992 and references therein.

\refis{BH} B. Halperin, {\it Helv. Phys. Acta} {\bf 56} 75 (1983)

\refis{wil}
The possibility of fractional Josephson frequency
appearing in single layered quantum Hall systems was
discussed by F. Wilczek in an unpublished talk at the
Conference on Fundamental Aspects of Quantum Theory,
Columbus, South Carolina, Dec. 1992. For
details, see F. Wilczek (in preparation).

\refis{cg} P.W. Anderson and G. Yuval, \journal J. Phys. C, 4, 607, 1971;
A. Schmid, \prl 51, 1506, 1983; F. Guinea, \pr B32, 7518, 1985.

\refis{MAC}
M. Rasolt and A.H. MacDonald, \pr 34, 5530, 1986;
H.A. Fertig, \prb 40, 1087, 1989;
A.H. MacDonald, P.M. Platzmann, and G.S. Boebinger, \prl 65, 775, 1990;
L. Brey, \prl 65, 903, 1990.

\refis{exp}
G.S. Boebinger {\it et\ al.,\/}
 \prl 64,  1793, 1990; Y.W. Suen {\it
et\ al.,\/} {\it ibid} {\bf 68}, 1379 (1992);
 J.P. Eisenstein {\it et\ al.,\/} {\it
ibid} {\bf 68}, 1383 (1992);
J.P.Eisenstein, L.N. Peiffer, and K.W. West, \prl 69, 3804, 1992.

\refis{JS} B.D. Josephson, \prl 1, 251, 1962.

\refis{edge} X.G. Wen, \journal Int. J. Mod. Phys., B6, 1711, 1992.

\endreferences
\vfill\eject

\subhead {\bf{Figure Captions}}

\item [Fig. 1] Sideways tunnelling junction . The tunnelling currents flow
in opposite directions in the first and second layers.

\end